\documentclass[prd,aps,groupedaddress,showpacs]{revtex4}
\usepackage{epsf,epsfig,graphicx}

\newcounter{muni}

\begin{document}
\hbadness=10000 \pagenumbering{arabic}

\title{Penguin-dominated $B\to PV$ decays in NLO perturbative QCD }

\author{Hsiang-nan Li$^{1}$}
\email{hnli@phys.sinica.edu.tw}
\author{Satoshi Mishima$^2$}
\email{mishima@ias.edu}

\affiliation{$^{1}$Institute of Physics, Academia Sinica, Taipei,
Taiwan 115, Republic of China,} \affiliation{$^{1}$Department of
Physics, National Cheng-Kung University, Tainan, Taiwan 701,
Republic of China}

\affiliation{$^{2}$School of Natural Sciences, Institute for
Advanced Study, Princeton, NJ 08540, U.S.A.}

\begin{abstract}

We study the penguin-dominated $B\to PV$ decays, with $P$ ($V$)
representing a pseudo-scalar (vector) meson, in the
next-to-leading-order (NLO) perturbative QCD (PQCD) formalism,
concentrating on the $B\to K\phi$, $\pi K^*$, $\rho K$, and
$\omega K$ modes. It is found that the NLO corrections
dramatically enhance the $B\to\rho K$, $\omega K$ branching
ratios, which were estimated to be small under the naive
factorization assumption. The patterns of the direct CP
asymmetries $A_{CP}(B^0\to\rho^\mp K^\pm)\approx
A_{CP}(B^\pm\to\rho^0 K^\pm)$ and $|A_{CP}(B^0\to\pi^\mp
K^{*\pm})|> |A_{CP}(B^\pm\to\pi^0 K^{*\pm})|$ are predicted,
differing from $|A_{CP}(B^0\to\pi^\mp K^\pm)|\gg
|A_{CP}(B^\pm\to\pi^0 K^\pm)|$. The above patterns, if confirmed
by data, will support the source of strong phases from the scalar
penguin annihilation in PQCD. The results for the mixing-induced
CP asymmetries $S_f$ are consistent with those obtained in the
literature, except that our $S_{\rho^0 K_S}$ is as low as 0.5.

\end{abstract}

\pacs{13.25.Hw, 12.38.Bx, 11.10.Hi}

\maketitle

\section{INTRODUCTION}

It has been a challenge to explain the branching ratios of the
penguin-dominated $B\to PV$ decays, such as $B\to K\phi$, $\pi
K^*$, $\rho K$, and $\omega K$. The predictions from the naive
factorization assumption (FA) \cite{BSW} and from the QCD-improved
factorization (QCDF) approach \cite{BBNS} are usually smaller than
the observed values \cite{fac,HMW,CY}. In order to account for the
data in QCDF, the hadronic parameters associated with the penguin
annihilation amplitudes must be tuned (i.e., adopting the scenario
``S4''), so that these amplitudes are large and constructive to
the leading FA ones \cite{BN}. On the contrary, the $B\to K\phi$
branching ratios have been predicted correctly
\cite{CKL,Mishima:2001ms} in the perturbative QCD (PQCD) approach
\cite{KLS,LUY}. The reason is that the leading amplitudes are not
factorizable in QCDF based on collinear factorization theorem due
to the existence of end-point singularities. Hence, they are
characterized by a scale of order $m_B$, $m_B$ being the $B$ meson
mass. The characteristic scale of the leading amplitudes in PQCD,
being factorizable since they are formulated in $k_T$
factorization theorem, is of order $\sqrt{m_B\Lambda}$, where
$\Lambda$ is a hadronic scale. At this lower characteristic scale,
the Wilson coefficients of the QCD penguin operators are enhanced,
so that the PQCD predictions for the $B\to K\phi$ branching ratios
are large enough. This mechanism has been called as the dynamical
enhancement of penguin contributions \cite{KLS}.

The PQCD analysis of the $B\to K\phi$ decays in
\cite{CKL,Mishima:2001ms} was performed at leading order (LO).
Recently, the PQCD formalism has been extended to
next-to-leading-order (NLO) accuracy \cite{LMS05}, where the
contributions from the NLO running of the Wilson coefficients, the
vertex corrections, the quark loops, and the magnetic penguin were
taken into account. It was found that these NLO contributions
affect penguin-dominated decays more than tree-dominated ones. For
example, the quark-loop and magnetic-penguin corrections decrease
the penguin amplitudes by about 10\%, and the $B\to\pi K$
branching ratios by about 20\% \cite{LMS05}. However, these NLO
contributions have almost no impact on the tree-dominated
$B\to\pi\pi$ branching ratios. Therefore, it is essential to
examine the NLO effects on LO predictions for other
penguin-dominated modes. In this paper we shall study the
penguin-dominated $B\to PV$ decays in NLO PQCD, concentrating on
$B\to K\phi$, $\pi K^*$, $\rho K$, and $\omega K$. It is expected
that the LO PQCD predictions for the tree-dominated modes, such as
$B\to\rho\pi$, $\omega\pi$ \cite{LMY}, are stable under the above
NLO corrections.

The second motivation to study the $B\to PV$ decays arises from
the observed direct CP asymmetries of the $B^\pm\to \rho^0K^\pm$
decays: the data of $A_{CP}(B^\pm\to \pi^0K^\pm)$ almost vanish,
but those of $A_{CP}(B^\pm\to \rho^0K^\pm)$ are sizable
\cite{HFAG}. The direct CP asymmetries of both the charged $B$
meson decays were predicted to be large in LO PQCD
\cite{KLS,Chen01}. However, as demonstrated in \cite{LMS05}, the
aforementioned NLO corrections can lower the LO predictions for
$|A_{CP}(B^\pm\to\pi^0 K^\pm)|$ to the measured values. The
responsible mechanism comes from the vertex corrections, which
induce a large and almost imaginary color-suppressed tree
amplitude. This amplitude then decreases the relative strong phase
between the total tree amplitude and the total penguin amplitude
involved in the $B^\pm\to\pi^0 K^\pm$ modes. Hence, it is natural
to investigate whether the LO predictions for
$A_{CP}(B^\pm\to\rho^0 K^\pm)$ and for the direct CP asymmetry of
another charged mode, $A_{CP}(B^\pm\to \pi^0K^{*\pm})$, will be
modified by the vertex corrections significantly.

The third motivation concerns the extraction of the standard-model
parameter $\sin(2\phi_1)$, $\phi_1$ being the weak phase, from the
mixing-induced CP asymmetries $S_f$ of the penguin-dominated $B\to
f$ decays. In principle, the results should be identical to those
from the tree-dominated $b\to c\bar c s$ transitions, such as
$B\to J/\psi K^{(*)}$. Nevertheless, potentially substantial
deviations $\Delta S\equiv S_{\rm penguin}-S_{c\bar cs}<0$ between
the above two extractions have been observed \cite{HFAG}.
Theoretical estimates
\cite{B05,BHNR,CCS2,LMS05,X,MS05,BPS05,WZ0610,Chiang03,GRZ} of the
tree pollution in the penguin-dominated decays are then crucial
for justifying these discrepancies as promising new physics
signals, which, however, gave tiny and positive $\Delta S$ for
most modes so far. The mixing-induced CP asymmetry of the
$B^0\to\pi^0 K_S$ decay has been calculated in NLO PQCD
\cite{LMS05}, and the result is consistent with those obtained in
the literature. It is necessary to complete the survey by
computing the mixing-induced CP asymmetries of the
penguin-dominated $B\to PV$ decays. Since the deviations in the
penguin-dominated modes are mainly caused by the color-suppressed
tree amplitude, which is sensitive to the vertex corrections, we
shall adopt the NLO PQCD formalism here.

It will be demonstrated that the dynamical enhancement of the
penguin amplitudes in the $B\to PV$ decays sustains the NLO
corrections. Especially, the NLO effects dramatically enhance the
$B\to\rho K$, $\omega K$ branching ratios, which were predicted to
be small in LO PQCD \cite{Chen01} and in FA \cite{fac}. The NLO
PQCD results for all the considered $B\to PV$ branching ratios
then match the data within theoretical uncertainties. Note that
the measured $B\to\omega K$ branching ratios are close to the
$B\to\rho K$ ones, though the former receive the additional
contribution from the Wilson coefficient $a_5$. Because the
magnitude of $a_5$ increases rapidly at a low energy scale, the
similarity between these two modes implies that two-body
nonleptonic $B$ meson decays are insensitive to low-energy
dynamics. Their data, if persisting, will encourage the
application of a perturbation theory to these processes.

The LO PQCD prediction for the direct CP asymmetry
$A_{CP}(B^\pm\to \rho^0K^\pm)$ \cite{Chen01} turns out to be
stable under the NLO corrections. That is, we predict the
different patterns $A_{CP}(B^0\to\rho^\mp K^\pm)\approx
A_{CP}(B^\pm\to\rho^0 K^\pm)$ and $|A_{CP}(B^0\to\pi^\mp
K^\pm)|\gg |A_{CP}(B^\pm\to\pi^0 K^\pm)|$, which are attributed to
the different orientations of the complex penguin amplitudes: the
penguin amplitude is inclined to the imaginary (real) axis in the
former (latter). When the penguin amplitude is almost imaginary,
its strong phase relative to the tree amplitude is maximal. The
color-suppressed tree amplitude, despite of being enhanced by the
vertex corrections, does not change it much. The configuration of
the relevant amplitudes in the $B\to\pi K^*$ decays is somewhat
located between the $B\to\rho K$ and $B\to\pi K$ ones, giving
$|A_{CP}(B^0\to\pi^\mp K^{*\pm})|> |A_{CP}(B^\pm\to\pi^0
K^{*\pm})|$. We point out that the above patterns can not be
attained by the other theoretical approaches, including QCDF
\cite{BBNS,BN}, soft-collinear effective theory (SCET)
\cite{BPS05,WZ0610,BPRS}, and QCDF plus final-state interactions
(FSI) \cite{CCS}.

As to the mixing-induced CP asymmetries, we obtain $\Delta S>0$ at
LO for all the $B\to PV$ modes. The NLO corrections increase most
of $\Delta S$ except that for $B^0\to\rho^0 K_S$, whose
mixing-induced CP asymmetry becomes as low as 0.5. The uniqueness
of the $B^0\to\rho^0 K_S$ decay results from the fact that both
the penguin amplitude and the color-suppressed tree amplitude are
almost imaginary and parallel to each other at NLO, a
configuration which leads to the very negative $\Delta S$. Among
the penguin-dominated decays we have investigated ($B^0\to\pi^0
K_S$, $\phi K_S$, $\rho^0 K_S$, $\omega K_S$), $B^0\to\phi K_S$ is
the cleanest one with the minimal tree pollution for extracting
$\sin(2\phi_1)$. Roughly speaking, the NLO PQCD predictions for
the $B\to PV$ mixing-induced CP asymmetries are consistent with
those in QCDF \cite{B05,BHNR}, but differ from those in QCDF plus
FSI \cite{CCS2}, which gave $\Delta S \approx +0.04$ for the
$B^0\to\rho^0 K_S$ decay \footnote{Notice $\Delta S \approx -0.04$
in \cite{CKC}.}.

We review the LO PQCD factorization formulas for the $B\to PV$
decays, and derive the NLO ones in Sec.~II. The branching ratios,
the direct CP asymmetries, and the mixing-induced CP asymmetries
of the penguin-dominated $B\to PV$ modes are calculated using the
NLO PQCD formalism in Sec.~III. We have updated the meson
distribution amplitudes and the relevant standard-model inputs in
the calculation. Section IV is the conclusion.

\section{FACTORIZATION FORMULAS}

The amplitude for a $B$ meson decay into the two-body final state
$M_2M_3$ through the $\bar b\to \bar s$ transition has the general
expression,
\begin{eqnarray}
{\cal A}(B \to M_2M_3) &=& V_{ub}^*V_{us}\, {\cal
A}^{(u)}_{M_2M_3}
 +V_{cb}^*V_{cs}\, {\cal A}^{(c)}_{M_2M_3}
 +V_{tb}^*V_{ts}\, {\cal A}^{(t)}_{M_2M_3}\;.
\label{eq:amp}
\end{eqnarray}
The amplitudes ${\cal A}^{(u)}_{M_2M_3}$, ${\cal
A}^{(c)}_{M_2M_3}$, and ${\cal A}^{(t)}_{M_2M_3}$ are decomposed
at LO into
\begin{eqnarray}
{\cal A}^{(u)}_{M_2M_3} &=& f_{M_3} F_e+{\cal M}_e+f_{M_2}
F_{eM_3}+{\cal M}_{eM_3} + f_BF_a+{\cal M}_a \;,
\nonumber\\
{\cal A}^{(c)}_{M_2M_3} &=& 0 \;,
\nonumber\\
{\cal A}^{(t)}_{M_2M_3} &=& - \left( f_{M_3} F^P_e+{\cal M}^P_e
+f_{M_2} F^P_{eM_3}
        +{\cal M}^P_{eM_3} +f_BF^P_a+{\cal M}^P_a \right)
\;,\label{decom}
\end{eqnarray}
where $f_B$ ($f_{M_2}$, $f_{M_3}$) is the $B$ meson ($M_2$ meson,
$M_3$ meson) decay constant, $F_e$ (${\cal M}_{e}$) the
color-allowed factorizable (nonfactorizable) tree emission
contribution, $F_{eM_3}$ (${\cal M}_{eM_3}$) the color-suppressed
factorizable (nonfactorizable) tree emission contribution, $F_a$
(${\cal M}_{a}$) the factorizable (nonfactorizable) tree
annihilation contribution, and those with the additional
superscripts $P$ the contributions from the penguin operators.
With the amplitude ${\cal A}(B\to M_2M_3)$ in Eq.~(\ref{eq:amp}),
we derive the branching ratios, the direct CP asymmetries, and the
mixing-induced CP asymmetries of two-body nonleptonic $B$ meson
decays, which have been defined in \cite{LMS05}.

\begin{table}[hb]
\begin{center}
\begin{tabular}{l|cc}
\hline\hline & ${\cal A}_{K^{+}\phi}^{(u)} $ & ${\cal
A}_{K^{0}\phi}^{(u)}$
\\\hline
$F_e$ & 0 &0
\\
${\cal M}_{e}$ & 0 & 0
\\
$F_a$ & $F_{a4} \left( a_1 \right)$ & 0
\\
${\cal M}_{a}$ & ${\cal M}_{a4} \left( a_1'\right)$ & 0
\\\hline
& ${\cal A}_{K^{+}\phi}^{(t)} $ & ${\cal A}_{K^{0}\phi}^{(t)}$
\\\hline
$F_e^P$ & $F_{e4} \left( a_3^{(s)}+ a_4^{(s)}+a_5^{(s)} \right)$ &
$F_{e4} \left( a_3^{(s)}+ a_4^{(s)}+a_5^{(s)} \right)$
\\
${\cal M}_{e}^P$ & ${\cal M}_{e4}\left(  a_3^{\prime (s)}
                                       + a_4^{\prime (s)}
- a_5^{\prime (s)}  \right)
                 +  {\cal M}_{e6}\left( a_6^{\prime (s)}  \right)$ &
${\cal M}_{e4}\left(  a_3^{\prime (s)}
                                       + a_4^{\prime (s)}
- a_5^{\prime (s)}  \right)
                 +  {\cal M}_{e6}\left( a_6^{\prime (s)}  \right)$
\\
$F_a^P$ & $F_{a4} \left( a_4^{(u)} \right)
              + F_{a6} \left( a_6^{(u)} \right)$ &
$F_{a4} \left( a_4^{(d)} \right)
              + F_{a6} \left( a_6^{(d)} \right)$
\\
${\cal M}_{a}^P$ & ${\cal M}_{a4}\left( a_4^{\prime (u)} \right)
                + {\cal M}_{a6}\left( a_6 ^{\prime (u)} \right)$ &
${\cal M}_{a4}\left( a_4^{\prime (d)} \right)
                + {\cal M}_{a6}\left( a_6 ^{\prime (d)} \right)$
\\
\hline\hline
\end{tabular}
\caption{$B\to K \phi$ decay amplitudes in LO PQCD.
}\label{kphi}
\end{center}
\end{table}

\begin{table}[hb]
\begin{center}
\begin{tabular}{l|cc}
\hline\hline & ${\cal A}_{\pi^+ K^{*0}}^{(u)} $ & $\sqrt{2}{\cal
A}_{\pi^0 K^{*+}}^{(u)} $
\\\hline
$F_e$ & 0 & $F_{e4} \left( a_1  \right)$
\\
${\cal M}_{e}$ & 0 & ${\cal M}_{e4} \left(  a_1' \right)$
\\
$F_{eK^*}$ & 0 & $F_{eK^*4} \left( a_2 \right)$
\\
${\cal M}_{eK^*}$ & 0 & ${\cal M}_{eK^*4}\left( a_2' \right)$
\\
$F_a$ & $F_{a4} \left( a_1 \right)$ & $F_{a4} \left( a_1 \right)$
\\
${\cal M}_{a}$ & ${\cal M}_{a4} \left( a_1'\right)$ & ${\cal
M}_{a4} \left( a_1'\right)$
\\\hline
& ${\cal A}_{\pi^+ K^{*0}}^{(t)} $ & $\sqrt{2}{\cal A}_{\pi^0
K^{*+}}^{(t)} $
\\\hline
$F_e^P$ & $F_{e4} \left( a_4^{(d)} \right)$   & $F_{e4} \left(
a_4^{(u)} \right)$
\\
${\cal M}_{e}^P$ & ${\cal M}_{e4}\left(  a_4^{\prime (d)} \right)
  + {\cal M}_{e6}\left( a_6^{\prime (d)} \right)$ &
${\cal M}_{e4}\left(  a_4^{\prime (u)} \right)
  + {\cal M}_{e6}\left( a_6^{\prime (u)} \right)$
\\
$F_{eK^*}^P$ & 0 & $F_{eK^*4} \left( a_3^{(u)} - a_3^{(d)} -
a_5^{(u)} + a_5^{(d)} \right)$
\\
${\cal M}_{eK^*}^P$ & 0 & ${\cal M}_{eK^*4}\left( a_3^{\prime (u)}
- a_3^{\prime (d)}
+ a_5^{\prime (u)} - a_5^{\prime (d)}  \right)$
\\
$F_a^P$ & $F_{a4} \left( a_4^{(u)} \right) + F_{a6} \left(
a_6^{(u)} \right)$ & $F_{a4} \left( a_4^{(u)} \right) + F_{a6}
\left( a_6^{(u)} \right)$
\\
${\cal M}_{a}^P$ & ${\cal M}_{a4}\left( a_4^{\prime (u)} \right)
  + {\cal M}_{a6}\left( a_6 ^{\prime (u)} \right)$ &
${\cal M}_{a4}\left( a_4^{\prime (u)} \right)
  + {\cal M}_{a6}\left( a_6 ^{\prime (u)} \right)$
\\
\hline & ${\cal A}_{\pi^- K^{*+}}^{(u)} $ & $\sqrt{2}{\cal
A}_{\pi^0 K^{*0}}^{(u)}$
\\\hline
$F_e$ & $F_{e4} \left( a_1  \right)$ & 0
\\
${\cal M}_{e}$ & ${\cal M}_{e4} \left(  a_1' \right)$ & 0
\\
$F_{eK^*}$ & 0 & $F_{eK^*4} \left( a_2 \right)$
\\
${\cal M}_{eK^*}$ & 0 & ${\cal M}_{eK^*4}\left( a_2' \right)$
\\
$F_a$ & 0 & 0
\\
${\cal M}_{a}$ & 0 & 0
\\\hline
& ${\cal A}_{\pi^- K^{*+}}^{(t)} $ & $\sqrt{2}{\cal A}_{\pi^0
K^{*0}}^{(t)} $
\\\hline
$F_e^P$ & $F_{e4} \left( a_4^{(u)} \right)$   & $F_{e4} \left( -
a_4^{(d)} \right)$
\\
${\cal M}_{e}^P$ & ${\cal M}_{e4}\left(  a_4^{\prime (u)} \right)
  + {\cal M}_{e6}\left( a_6^{\prime (u)} \right)$ &
${\cal M}_{e4}\left(  - a_4^{\prime (d)} \right)
   + {\cal M}_{e6}\left( - a_6^{\prime (d)}  \right)$
\\
$F_{eK^*}^P$ & 0 & $F_{eK^*4} \left( a_3^{(u)} - a_3^{(d)} -
a_5^{(u)} + a_5^{(d)} \right)$
\\
${\cal M}_{eK^*}^P$ & 0 & ${\cal M}_{eK^*4}\left( a_3^{\prime (u)}
- a_3^{\prime (d)}
+ a_5^{\prime (u)} - a_5^{\prime (d)}  \right)$
\\
$F_a^P$ & $F_{a4} \left( a_4^{(d)} \right) + F_{a6} \left(
a_6^{(d)} \right)$ & $F_{a4} \left( - a_4^{(d)} \right) + F_{a6}
\left( - a_6^{(d)} \right)$
\\
${\cal M}_{a}^P$ & ${\cal M}_{a4}\left( a_4^{\prime (d)} \right)
   + {\cal M}_{a6}\left( a_6 ^{\prime (d)} \right)$ &
${\cal M}_{a4}\left( - a_4^{\prime (d)} \right)
   + {\cal M}_{a6}\left( - a_6 ^{\prime (d)} \right)$
\\\hline\hline
\end{tabular}
\caption{$B \to \pi K^{*}$ decay amplitudes in LO PQCD.}
\label{pik}
\end{center}
\end{table}

\begin{table}[hb]
\begin{center}
\begin{tabular}{l|cc}
\hline\hline & ${\cal A}_{\rho^+ K^{0}}^{(u)} $ & $\sqrt{2}{\cal
A}_{\rho^0 K^{+}, \omega K^{+}}^{(u)} $
\\\hline
$F_e$ & 0 & $F_{e4} \left( a_1  \right)$
\\
${\cal M}_{e}$ & 0 & ${\cal M}_{e4} \left(  a_1' \right)$
\\
$F_{eK}$ & 0 & $F_{eK4} \left( a_2 \right)$
\\
${\cal M}_{eK}$ & 0 & ${\cal M}_{eK4}\left( a_2' \right)$
\\
$F_a$ & $F_{a4} \left( a_1 \right)$ & $F_{a4} \left( a_1 \right)$
\\
${\cal M}_{a}$ & ${\cal M}_{a4} \left( a_1'\right)$ & ${\cal
M}_{a4} \left( a_1'\right)$
\\\hline
& ${\cal A}_{\rho^+ K^{0}}^{(t)} $ & $\sqrt{2}{\cal A}_{\rho^0
K^{+},\omega K^+}^{(t)} $
\\\hline
$F_e^P$ & $F_{e4} \left( a_4^{(d)} \right) +  F_{e6} \left(
a_6^{(d)} \right)$ & $F_{e4} \left( a_4^{(u)} \right) + F_{e6}
\left( a_6^{(u)}  \right)$
\\
${\cal M}_{e}^P$ & ${\cal M}_{e4}\left(  a_4^{\prime (d)} \right)
  + {\cal M}_{e6}\left( a_6^{\prime (d)} \right)$ &
${\cal M}_{e4}\left(  a_4^{\prime (u)} \right)
  + {\cal M}_{e6}\left( a_6^{\prime (u)} \right)$
\\
$F_{eK}^P$ & 0 & $F_{eK4} \left( a_3^{(u)} \mp a_3^{(d)} +
a_5^{(u)} \mp a_5^{(d)} \right)$
\\
${\cal M}_{eK}^P$ & 0 & ${\cal M}_{eK4}\left( a_3^{\prime (u)} \mp
a_3^{\prime (d)}
- a_5^{\prime (u)} \pm a_5^{\prime (d)}  \right)$
\\
$F_a^P$ & $F_{a4} \left( a_4^{(u)} \right) + F_{a6} \left(
a_6^{(u)} \right)$ & $F_{a4} \left( a_4^{(u)} \right) + F_{a6}
\left( a_6^{(u)} \right)$
\\
${\cal M}_{a}^P$ & ${\cal M}_{a4}\left( a_4^{\prime (u)} \right)
  + {\cal M}_{a6}\left( a_6 ^{\prime (u)} \right)$ &
${\cal M}_{a4}\left( a_4^{\prime (u)} \right)
  + {\cal M}_{a6}\left( a_6 ^{\prime (u)} \right)$
\\
\hline & ${\cal A}_{\rho^- K^{+}}^{(u)} $ & $\sqrt{2}{\cal
A}_{\rho^0 K^{0},\omega K^0}^{(u)}$
\\\hline
$F_e$ & $F_{e4} \left( a_1  \right)$ & 0
\\
${\cal M}_{e}$ & ${\cal M}_{e4} \left(  a_1' \right)$ & 0
\\
$F_{eK}$ & 0 & $F_{eK4} \left( a_2 \right)$
\\
${\cal M}_{eK}$ & 0 & ${\cal M}_{eK4}\left( a_2' \right)$
\\
$F_a$ & 0 & 0
\\
${\cal M}_{a}$ & 0 & 0
\\\hline
& ${\cal A}_{\rho^- K^{+}}^{(t)} $ & $\sqrt{2}{\cal A}_{\rho^0
K^{0}, \omega K^0}^{(t)} $
\\\hline
$F_e^P$ & $F_{e4} \left( a_4^{(u)} \right) + F_{e6} \left(
a_6^{(u)} \right)$ & $F_{e4} \left( \mp a_4^{(d)} \right) +F_{e6}
\left( \mp a_6^{(d)} \right)$
\\
${\cal M}_{e}^P$ & ${\cal M}_{e4}\left(  a_4^{\prime (u)} \right)
  + {\cal M}_{e6}\left( a_6^{\prime (u)} \right)$ &
${\cal M}_{e4}\left(  \mp a_4^{\prime (d)} \right)
   + {\cal M}_{e6}\left( \mp a_6^{\prime (d)}  \right)$
\\
$F_{eK}^P$ & 0 & $F_{eK4} \left( a_3^{(u)} \mp a_3^{(d)} +
a_5^{(u)} \mp a_5^{(d)} \right)$
\\
${\cal M}_{eK}^P$ & 0 & ${\cal M}_{eK4}\left( a_3^{\prime (u)} \mp
a_3^{\prime (d)}
- a_5^{\prime (u)} \pm a_5^{\prime (d)}  \right)$
\\
$F_a^P$ & $F_{a4} \left( a_4^{(d)} \right) + F_{a6} \left(
a_6^{(d)} \right)$ & $F_{a4} \left( \mp a_4^{(d)} \right) + F_{a6}
\left( \mp a_6^{(d)} \right)$
\\
${\cal M}_{a}^P$ & ${\cal M}_{a4}\left( a_4^{\prime (d)} \right)
   + {\cal M}_{a6}\left( a_6 ^{\prime (d)} \right)$ &
${\cal M}_{a4}\left( \mp a_4^{\prime (d)} \right)
   + {\cal M}_{a6}\left( \mp a_6 ^{\prime (d)} \right)$
\\\hline\hline
\end{tabular}
\caption{$B \to \rho K$, $\omega K$ decay amplitudes in LO PQCD
with the upper (lower) signs applying to the $\rho K$ ($\omega K$)
modes.} \label{rhok}
\end{center}
\end{table}

The LO PQCD factorization formulas for the $B\to K\phi$, $\pi
K^*$, and $\rho(\omega) K$ decay amplitudes are summarized in
Tables~\ref{kphi}, \ref{pik}, and \ref{rhok}, respectively. The
Wilson coefficients $a^{(q)}$ for the factorizable contributions
and $a^{\prime(q)}$ for the nonfactorizable contributions are the
combinations of $a_i$ in the standard definitions \cite{LMS05},
where $q=u$, $d$ or $s$ denotes the quark pair produced in the
electroweak penguin. The explicit expression for each amplitude in
Tables~\ref{kphi}--\ref{rhok} is similar to that of the $B\to
P_2P_3$ modes \cite{LMS05}, but with the replacements,
\begin{eqnarray}
\phi^A_{3}  \to\  \phi_{3} \;,\ \ \ \ \phi^P_{3} \to\ \phi^s_{3}
\;,\ \ \ \ \phi^T_{3}  \to\  \phi^t_{3} \;, \ \ \ \ m_{02} \to\
m_{02}\;,\ \ \ \  m_{03}   \to\ -m_3\;,\label{re1}
\end{eqnarray}
for $B\to P_2 V_3$ (involving the $B\to P_2$ transition), and
\begin{eqnarray}
\phi^A_{2} \to\  \phi_{2} \;,\ \ \ \ \phi^P_{2}  \to\ \phi^s_{2}
\;,\ \ \ \ \phi^T_{2} \to\  \phi^t_{2} \;,\ \ \ \ m_{02}   \to\
-m_2\;,\ \ \ \  m_{03}   \to\ -m_{03}\;,\label{re2}
\end{eqnarray}
for $B\to V_2 P_3$ (involving the $B\to V_2$ transition), where
$m_0$ is the chiral enhancement scale associated with a
pseudo-scalar meson, and $m$ the vector meson mass. The above
replacements also apply to the factorization formulas for the NLO
corrections from the quark loops and from the magnetic penguin.

We add the various NLO corrections to the LO factorization
formulas in Tables~\ref{kphi}--\ref{rhok}. The NLO corrections to
the smaller nonfactorizable contributions will be neglected. It is
understood that the Wilson coefficients appearing below refer to
the NLO ones. The vertex corrections to the $B\to PV$ decays
modify the Wilson coefficients for the emission amplitudes in the
standard definition into
\begin{eqnarray}
a_1(\mu) &\to& a_1(\mu)
+\frac{\alpha_s(\mu)}{4\pi}C_F\frac{C_{1}(\mu)}{N_c} V_1(M) \;,
\nonumber\\
a_2(\mu) &\to& a_2(\mu)
+\frac{\alpha_s(\mu)}{4\pi}C_F\frac{C_{2}(\mu)}{N_c} V_2(M) \;,
\nonumber\\
a_i(\mu) &\to& a_i(\mu)
+\frac{\alpha_s(\mu)}{4\pi}C_F\frac{C_{i\pm 1}(\mu)}{N_c} V_i(M)
\;,\;\;\;\;i=3 - 10\;,\label{wnlo}
\end{eqnarray}
where $M$ denotes the meson emitted from the weak vertex, and the
upper (lower) sign applies for odd (even) $i$. When $M$ is a
pseudo-scalar meson, the functions $V_i(M)$ are given, in the NDR
scheme, by \cite{BBNS,BN}
\begin{eqnarray}
V_i(M) &=& \left\{ {\renewcommand\arraystretch{2.5}
\begin{array}{ll}
12\ln\displaystyle{\frac{m_b}{\mu}}-18
+\frac{2\sqrt{2N_c}}{f_M}\int_0^1 dx\, \phi_M^A(x)\, g(x)\;, &
\mbox{\rm for }i=1-4,9,10\;,
\\
-12\ln\displaystyle{\frac{m_b}{\mu}}+6
-\frac{2\sqrt{2N_c}}{f_M}\int_0^1dx\, \phi_M^A(x)\, g(1-x)\;, &
\mbox{\rm for }i=5,7\;,
\\
\displaystyle{ -6 +\frac{2\sqrt{2N_c}}{f_M}\int_0^1 dx\,
\phi_M^{P}(x)\, h(x) }\;, & \mbox{\rm for }i=6,8\;,
\end{array}
} \right.\label{vim}
\end{eqnarray}
$x$ being a momentum fraction. For a vector meson $M$, $\phi_M^A$
($\phi_M^P$) is replaced by $\phi_M$ ($-\phi_M^s$), and $f_M$ by
$f_M^T$ in the third line of the above formulas. The explicit
expressions of the hard kernels $g$ and $h$ can be found in
\cite{BN}.

When a vector meson is emitted from the weak vertex in, for
example, the cases of $B\to K\phi$, $\pi K^*$, the leading
emission contributions from the Wilson coefficient $a_6^{(q)}$ are
absent as indicated in Tables~\ref{kphi} and \ref{pik}. Including
the vertex corrections, the NLO piece $a^{(q)}_{6\rm VC}$,
containing the second terms of $a_{6,8}$ in Eq.~(\ref{wnlo}),
contributes through the following additional amplitudes,
\begin{eqnarray}
&K^{+,0}\phi: &f_\phi F^{P}_e\to f_\phi F^{P}_e+f_\phi^T
F_{e6} \left( a_{6\rm VC}^{(s)} \right)\;,\nonumber\\
&\pi^+K^{*0}: &f_{K^*} F^{P}_e\to f_{K^*} F^{P}_e+f_{K^*}^T
F_{e6} \left( a_{6\rm VC}^{(d)} \right)\;,\nonumber\\
&\pi^{0,-} K^{*+}: &f_{K^*} F^{P}_e\to f_{K^*} F^{P}_e+f_{K^*}^T
F_{e6} \left( a_{6\rm VC}^{(u)} \right)\;,\nonumber\\
&\pi^0 K^{*0}: &f_{K^*} F^{P}_e\to f_{K^*} F^{P}_e+f_{K^*}^T
F_{e6} \left( -a_{6\rm VC}^{(d)} \right)\;.
\end{eqnarray}
These additional amplitudes have emerged in the NLO analysis of
the $B\to VV$ decays \cite{LM06}.

Taking into account the NLO contributions from the quark loops and
from the magnetic penguin, the decay amplitudes in
Eq.~(\ref{eq:amp}) are modified into
\begin{eqnarray}
{\renewcommand\arraystretch{2.0}
\begin{array}{ll}
\displaystyle {\cal A}^{(u,c)}_{K^{+,0}\phi } \, \to\, {\cal
A}^{(u,c)}_{K^{+,0}\phi }+{\cal M}^{(u,c)}_{K\phi}\;,
&\displaystyle {\cal A}^{(t)}_{K^{+,0}\phi } \, \to\, {\cal
A}^{(t)}_{K^{+,0}\phi }-{\cal M}^{(t)}_{K\phi}-{\cal
M}^{(g)}_{K\phi}\;,
\\
\displaystyle {\cal A}^{(u,c)}_{\pi^+K^{*0}, \pi^-K^{*+}} \, \to\,
{\cal A}^{(u,c)}_{\pi^+K^{*0}, \pi^-K^{*+}}+{\cal M}^{(u,c)}_{\pi
K^*}\;, &\displaystyle {\cal A}^{(t)}_{\pi^+K^{*0}, \pi^-K^{*+}}
\, \to\, {\cal A}^{(t)}_{\pi^+K^{*0}, \pi^-K^{*+}}-{\cal
M}^{(t)}_{\pi K^*}-{\cal M}^{(g)}_{\pi K^*}\;,
\\
\displaystyle {\cal A}^{(u,c)}_{\pi^0K^{*+} } \, \to\, {\cal
A}^{(u,c)}_{\pi^0K^{*+} } +\frac{1}{\sqrt{2}}{\cal M}^{(u,c)}_{\pi
K^*}\;, &\displaystyle
 {\cal A}^{(t)}_{\pi^0K^{*+} } \, \to\, {\cal A}^{(t)}_{\pi^0K^{*+} }
-\frac{1}{\sqrt{2}}{\cal M}^{(t)}_{\pi
K^*}-\frac{1}{\sqrt{2}}{\cal M}^{(g)}_{\pi K^*}\;,
\\
\displaystyle {\cal A}^{(u,c)}_{\pi^0K^{*0} } \, \to\, {\cal
A}^{(u,c)}_{\pi^0K^{*0} } -\frac{1}{\sqrt{2}}{\cal M}^{(u,c)}_{\pi
K^*}\;, & \displaystyle {\cal A}^{(t)}_{\pi^0K^{*0} } \, \to\,
{\cal A}^{(t)}_{\pi^0K^{*0} } +\frac{1}{\sqrt{2}}{\cal
M}^{(t)}_{\pi K^*}+\frac{1}{\sqrt{2}}{\cal M}^{(g)}_{\pi K^*}\;,
\\
\displaystyle {\cal A}^{(u,c)}_{\rho^+K^{0}, \rho^-K^{+}} \, \to\,
{\cal A}^{(u,c)}_{\rho^+K^{0}, \rho^-K^{+}}+{\cal M}^{(u,c)}_{\rho
K}\;, &\displaystyle {\cal A}^{(t)}_{\rho^+K^{0}, \rho^-K^{+}} \,
\to\, {\cal A}^{(t)}_{\rho^+K^{0}, \rho^-K^{+}}-{\cal
M}^{(t)}_{\rho K}-{\cal M}^{(g)}_{\rho K}\;,
\\
\displaystyle {\cal A}^{(u,c)}_{\rho^0K^{+}, \omega K^+ } \, \to\,
{\cal A}^{(u,c)}_{\rho^0K^{+}, \omega K^+}
+\frac{1}{\sqrt{2}}{\cal M}^{(u,c)}_{\rho K, \omega K}\;,
&\displaystyle
 {\cal A}^{(t)}_{\rho^0K^{+}, \omega K^+} \, \to\,
{\cal A}^{(t)}_{\rho^0K^{+}, \omega K^+} -\frac{1}{\sqrt{2}}{\cal
M}^{(t)}_{\rho K, \omega K}-\frac{1}{\sqrt{2}}{\cal M}^{(g)}_{\rho
K, \omega K}\;,
\\
\displaystyle {\cal A}^{(u,c)}_{\rho^0K^{0}, \omega K^{0} } \,
\to\, {\cal A}^{(u,c)}_{\rho^0K^{0}, \omega K^{0} }
\mp\frac{1}{\sqrt{2}}{\cal M}^{(u,c)}_{\rho K, \omega K}\;, &
\displaystyle {\cal A}^{(t)}_{\rho^0K^{0}, \omega K^{0} } \, \to\,
{\cal A}^{(t)}_{\rho^0K^{0}, \omega K^{0}}
\pm\frac{1}{\sqrt{2}}{\cal M}^{(t)}_{\rho K, \omega
K}\pm\frac{1}{\sqrt{2}}{\cal M}^{(g)}_{\rho K, \omega K}\;,
\end{array}
}
\end{eqnarray}
where the upper (lower) signs apply to the $\rho K$ ($\omega K$)
decays. The NLO amplitudes ${\cal M}_{M_2M_3}^{(u)}$, ${\cal
M}_{M_2M_3}^{(c)}$, ${\cal M}_{M_2M_3}^{(t)}$, and ${\cal
M}_{M_2M_3}^{(g)}$ denote the up-loop, charm-loop,
QCD-penguin-loop, and magnetic-penguin corrections, respectively.
Their expressions are similar to those for the $B\to PP$ decays
\cite{LMS05}, but with the replacements specified in
Eqs.~(\ref{re1}) and (\ref{re2}). The magnetic-penguin
contribution to the $B\to PV$ modes has been computed in
\cite{MISHIMA03}. It will be shown that the quark-loop corrections
are always constructive, increasing all the $B\to PV$ branching
ratios  considered here. The magnetic-penguin corrections decrease
the $B\to K\phi$, $\pi K^{*}$, $\omega K$ branching ratios, but
enhance the $B\to\rho K$ ones slightly.

\section{NUMERICAL RESULTS}

We perform the numerical analysis in this section based on the
factorization formulas derived above. The choices of the $B$ meson
wave function, the $B$ meson lifetimes, the decay constants and
the chiral enhancement scales associated with the pseudo-scale
mesons, and the weak phase $\phi_1=21.7^o$ are the same as in
\cite{LMS05}. We take $m_\rho=0.77$ GeV, $m_\omega=0.78$ GeV
$m_{K^*}=0.89$ GeV, and $m_{\phi}=1.02$ GeV for the vector meson
masses \cite{PDG}, and $m_b=4.8$ GeV for the $b$ quark mass
appearing in the magnetic-penguin operator $O_{8g}$. The
longitudinal decay constants of the vector mesons can be extracted
from other decay rates, for example, $f_{\rho,K^*}$ from the
measured $\tau^-\to(\rho^-,K^{*-})\nu_\tau$ decay rates
\cite{PDG,BZ0412}. The transverse decay constants have been
evaluated in QCD sum rules, which are summarized
below~\cite{BZ0412,BZ0603,BZ0510}:
\begin{eqnarray}
& &f_{\rho}=(0.209\pm 0.002)\; {\rm GeV}\;,\;\;\;\;
f_{\rho}^T=(0.165\pm 0.009)\; {\rm GeV}
\;,\nonumber\\
& &f_{\omega}=(0.195\pm 0.003)\; {\rm GeV}\;,\;\;\;\;
f_{\omega}^T=(0.145\pm 0.010)\; {\rm GeV}
\;,\nonumber\\
& &f_{K^*}=(0.217\pm 0.005)\; {\rm GeV}\;,\;\;
f_{K^*}^T=(0.185\pm 0.010)\; {\rm GeV}
\;,\nonumber\\
& &f_{\phi}=(0.231\pm 0.004)\; {\rm GeV}\;,\;\;\;\;
f_{\phi}^T=(0.200\pm 0.010)\; {\rm GeV}
\;.\label{mdc}
\end{eqnarray}

We employ the updated twist-2 and twist-3 pseudo-scalar meson
distribution amplitudes, and the updated twist-2 vector meson
distribution amplitudes from QCD sum rules \cite{BZ0603,BBL0603}.
For the twist-3 vector meson distribution amplitudes, we adopt the
asymptotic models \cite{Li04}, since some of them are still
uncertain in sum rules. The explicit expressions of the above
distribution amplitudes are collected in the Appendix. The
sum-rule analysis of the kaon and $K^*$ meson distribution
amplitudes have been also performed in \cite{KMM} and \cite{BL04},
respectively. When analyzing the theoretical uncertainty, we
consider the following ranges for the Gegenbauer coefficients in
the light-meson distribution amplitudes, whose definitions are
referred to the Appendix,
\begin{eqnarray}
& &a_2^\pi=0.25\pm 0.25\;,\;\;\;\; a_2^K=0.25\pm
0.25\;,\nonumber\\
& &a_{2\rho}^\parallel=a_{2\omega}^\parallel=
0.15\pm 0.15\;,\;\;\;\;a_{2K^*}^\parallel
=0.11\pm 0.11 \;,\;\;\;\;  a_{2\phi}^\parallel = 0.0\pm 0.2\;.
\label{gc}
\end{eqnarray}
Since our results do not vary with the Gegenbauer coefficients
$a_1^K$ and $a_{1K^*}^\parallel$ much, their values have been
fixed at $a_1^K=0.06$ and $a_{1K^*}^\parallel =0.03$. All the
Gegenbauer coefficients associated with the twist-3 distribution
amplitudes are also fixed for simplicity. We update the CKM matrix
elements \cite{PDG},
\begin{eqnarray}
& &V_{ud}=0.97377\;,\;\;\;\;V_{us}=0.2257\;,\;\;\;\;
|V_{ub}|=(4.31\pm 0.30)\times 10^{-3}\;,\nonumber\\
& &V_{cd}=-0.230\;,\;\;\;\;V_{cs}=0.957\;,\;\;\;\;
V_{cb}=0.0416\;,\nonumber\\
& &\phi_3 = (70\pm 30)^\circ\;.\label{ckm}
\end{eqnarray}
The increase of $|V_{ub}|$ from its old value $(3.67\pm
0.47)\times 10^{-3}$ will enhance the direct CP asymmetries of $B$
meson decays as shown later.

We then evaluate the relevant transition form factors at maximal
recoil,
\begin{eqnarray}
& &F_0^{B\pi} = 0.27^{+0.07}_{-0.05} \;,\;\;\;\; F_0^{BK} =
0.43^{+0.10}_{-0.08}
\;,\nonumber\\
& &A_0^{B\rho} = 0.32^{+0.07}_{-0.06} \;,\;\;\;\; A_0^{B\omega} =
0.29^{+0.06}_{-0.05} \;,\;\;\;\; A_0^{BK^*} = 0.38^{+0.07}_{-0.06}
\;,\label{form}
\end{eqnarray}
where the theoretical uncertainties come from the variation of the
parameters associated with the $B$ meson wave function, and from
Eqs.~(\ref{gc}) and (\ref{ckm}). The above values of $F_0^{B\pi}$
and $F_0^{BK}$ are higher than $F_0^{B\pi}=0.24$ and
$F_0^{BK}=0.36$ obtained in \cite{LMS05} from the old pseudo-scalar
meson distribution amplitudes. The others are similar to
$A_0^{B\rho}=0.30$, $A_0^{B\omega}=0.28$, and $A_0^{BK^*}=0.37$
from QCD sum rules \cite{BZ0412}. Our $A_0^{BK^*}$ is a bit above
that from the covariant light-front QCD calculation \cite{CCH}.
Note that larger values of the $B\to V$ form factors have been
adopted in QCDF \cite{BN} in order to account for the observed
$B\to PV$ branching ratios.

\begin{table}[hbt]
\begin{center}
\begin{tabular}{ccccc}
\hline\hline $\mu$ (GeV)&$0.5$&
$1.0$&
$1.5$&
$2.0$\\
\hline $a_1(\mu)$&$1.20$&
$1.10$&
$1.07$&
$1.05$
\\
$a_{1VC}(\mu)$&$-0.25 + i\, 0.29$&
$0.00 + i\, 0.09$&
$0.02 + i\, 0.06$&
$0.03 + i\, 0.04$
\\
\hline $a_2(\mu)$&$-0.26$&
$-0.06$&
$0.02$&
$0.07$
\\
$a_{2VC}(\mu)$&$0.47 - i\, 0.53$&
$0.00 - i\, 0.23$&
$-0.08 - i\, 0.17$&
$-0.12 - i\, 0.14$
\\
\hline $a_3(\mu)$&$0.03$&
$0.01$&
$0.01$&
$0.01$
\\
$a_{3VC}(\mu)$&$-0.06 + i\, 0.06$&
$0.00 + i\, 0.02$&
$0.00 + i\, 0.01$&
$0.01 + i\, 0.01$
\\
\hline $a_4(\mu)$&$-0.14$&
$-0.08$&
$-0.06$&
$-0.05$
\\
$a_{4VC}(\mu)$&$0.03 - i\, 0.04$&
$0.00 - i\, 0.01$&
$0.00 + i\, 0.00$&
$0.00 + i\, 0.00$
\\
\hline $a_5(\mu)$&$-0.15$&
$-0.04$&
$-0.02$&
$-0.02$
\\
$a_{5VC}(\mu)$&$0.29 - i\, 0.13$&
$0.03 - i\, 0.03$&
$0.01 - i\, 0.01$&
$0.00 - i\, 0.01$
\\
\hline $a_6(\mu)$&$-0.37$&
$-0.14$&
$-0.10$&
$-0.08$
\\
$a_{6VC}(\mu)$&$0.00 + i\, 0.00$&
$0.00 + i\, 0.00$&
$0.00 + i\, 0.00$&
$0.00 + i\, 0.00$
\\
\hline\hline
\end{tabular}
\end{center}
\caption{RG evolution of the Wilson coefficients involved in the
$B\to \omega K$ decays, where $a_i(\mu)$ and $a_{i\,VC}(\mu)$
represent the first and second terms on the right-hand side of
Eq.~(\ref{wnlo}), respectively.}\label{scale}
\end{table}

We present in Table~\ref{scale} the renormalization-group (RG)
evolution of the Wilson coefficients $a_i$ and of the vertex
corrections $a_{iVC}$ in the scale $\mu$ for the $B\to\omega K$
decays, which correspond to the first and second terms on the
right-hand side of Eq.~(\ref{wnlo}), respectively. The RG
evolution exhibited by the other $B\to PV$ modes is similar. It is
seen that the coefficients $a_{4-6}$ associated with the penguin
operators are enhanced more than the leading tree coefficient
$a_1$ at a low $\mu$. This mechanism, referred to the dynamical
enhancement in \cite{KLS}, gives the large branching ratios of
penguin-dominated $B$ meson decays \cite{KLS,CKL,Mishima:2001ms}.
Most of the vertex corrections remain subleading even at a low
$\mu$, except those to $a_2$ and $a_5$. Though the correction
$a_{3VC}$ to $a_3$ is significant, its effect, much smaller than
that of $a_{5VC}$, is negligible. The large $a_{2VC}$ has been
found to be responsible for the reduction of the LO PQCD
predictions for the direct CP asymmetry $|A_{CP}(B^\pm\to\pi^0
K^\pm)|$ \cite{LMS05} as explained in the Introduction. The new
effect observed in this work arises from the large $a_{5VC}$,
which will be elucidated below.

\begin{table}[hbt]
\begin{center}
\begin{tabular}{c|ll|ll|ll|ll}
\hline\hline & \multicolumn{2}{c|}{$K\phi$} &
\multicolumn{2}{c|}{$\pi K^*$} & \multicolumn{2}{c|}{$\rho K$} &
\multicolumn{2}{c}{$\omega K$}
\\
\hline &\ \ \ LO &\ \ +NLO &\ \ \ LO &\ \ +NLO &\ \ \ LO &\ \ +NLO
&\ \ \ LO &\ \ +NLO
\\
\hline $T'$ &
$\phantom{0}0.0\,e^{i\;0.0}$ & $\phantom{0}0.4\,e^{-i\;1.6}$ &
$12.3\,e^{i\;0.0}$ & $12.0\,e^{i\;0.1}$ &
$10.6\,e^{i\;0.0}$ & $10.3\,e^{i\;0.1}$ &
$\phantom{0}9.6\,e^{i\;0.0}$ & $\phantom{0}9.4\,e^{i\;0.1}$
\\
$C'$ &
$\phantom{0}0.0\,e^{i\;0.0}$ & $\phantom{0}0.4\,e^{i\;1.6}$ &
$\phantom{0}0.5\,e^{-i\;1.9}$ & $\phantom{0}2.4\,e^{-i\;1.2}$ &
$\phantom{0}0.8\,e^{-i\;2.3}$ & $\phantom{0}4.4\,e^{-i\;1.3}$ &
$\phantom{0}0.7\,e^{-i\;2.3}$ & $\phantom{0}4.1\,e^{-i\;1.3}$
\\
$P'$ &
$36.1\,e^{i\;2.8}$ & $41.2\,e^{i\;2.8}$ &
$21.1\,e^{i\;2.6}$ & $23.0\,e^{i\;2.6}$ &
$17.1\,e^{-i\;1.5}$& $25.2\,e^{-i\;1.4}$ &
$15.7\,e^{-i\;1.5}$& $23.2\,e^{-i\;1.4}$
\\
$P^{c\,\prime}$ &   $\phantom{0}0.9\,e^{-i\;2.6}$ &
$11.8\,e^{-i\;0.5}$
&\ \ \ \ ---&\ \ \ \ --- &\ \ \ \ ---&\ \ \ \
--- & $\phantom{0}1.8\,e^{-i\;2.5}$ & $19.8\,e^{-i\;0.5}$
\\
$P'_{ew}$ &
$11.1\,e^{i\;3.1}$& $11.4\,e^{-i\;3.1}$ &
$\phantom{0}5.8\,e^{i\;3.1}$ & $\phantom{0}5.6\,e^{-i\;3.0}$ &
$\phantom{0}9.9\,e^{i\;3.1}$ & $10.1\,e^{i\;3.1}$ &
$\phantom{0}3.1\,e^{i\;3.1}$ & $\phantom{0}3.2\,e^{i\;3.1}$
\\
\hline\hline
\end{tabular}
\end{center}
\caption{Topological amplitudes in units of $10^{-5}$ GeV.}
\label{ta}
\end{table}

Since all the meson distribution amplitudes derived from QCD sum
rules are defined at the scale 1 GeV, we propose to freeze the RG
evolution of the Wilson coefficients at this scale, when the
energy runs below it. That is, we set $C_i(\mu)=C_i(1\;{\rm GeV})$
as $\mu\le 1$ GeV. Note that the freezing scale was chosen as 0.5
GeV in \cite{LMS05,LM06}. We have checked that the new choice of
the freezing scale decreases the NLO PQCD predictions for the
$B\to\pi K$ branching ratios by about 15\%, and has a tiny effect
on those for the $B\to\pi\pi$ branching ratios. The final outcomes
are still within the theoretical uncertainty in \cite{LMS05,LM06}.
In order to facilitate the discussion, we present the central
values of our predictions in terms of the topological amplitudes
in Table~\ref{ta}, where $T'$, $C'$, $P'$, $P^{c\prime}$, and
$P'_{ew}$ denote the color-allowed tree, color-suppressed tree,
QCD penguin, color-suppressed QCD penguin, and electro-weak
penguin contributions, respectively. Their definitions are
referred to \cite{Charng}. Note that $P^{c\prime}$ exists only in
the $B\to K\phi$ and $B\to\omega K$ decays due to the
flavor-singlet quark content of the $\phi$ and $\omega$ mesons. We
have estimated the averaged hard scale $\langle t\rangle$ for each
of the above topological amplitude, and found $\langle
t\rangle\approx 2.0$ GeV for $T'$ and $P_{ew}'$, $\langle
t\rangle\approx 1.5-2.0$ GeV for $P'$, and $\langle
t\rangle\approx 1.0-1.5$ GeV for $C'$ and $P^{c\prime}$. That is,
the hard scales, despite of being all of $O(\sqrt{m_B\Lambda})$,
vary a bit among the different topological amplitudes.

\begin{table}[hbt]
\begin{center}
\begin{tabular}{cccccccccc}
\hline\hline Mode & Data~\cite{HFAG}& QCDF(S4)\,\cite{BN} &
 LO & LO$_{\rm NLOWC}$ & +VC
& +QL &  +MP  & +NLO
\\
\hline $B^\pm\to K^\pm \phi $ & $8.30\pm 0.65$ & 11.6 &
13.8&41.8&14.3&44.3&28.4& $7.8^{+5.9\,(+5.6)}_{-1.8\,(-1.7)}$
\\
$B^0\to K^0 \phi$ & $8.3^{+1.2}_{-1.0}$ & 10.5 &
12.9&39.2&13.4&41.5&26.6& $7.3^{+5.4\,(+5.1)}_{-1.6\,(-1.5)}$
\\
\hline $B^\pm\to \pi^\pm K^{*0}$ & $ 10.7 \pm 0.8 $ & 8.4 &
5.5&10.6&9.6&11.3&6.5& $6.0^{+2.8\,(+2.7)}_{-1.5\,(-1.4)}$
\\
$B^\pm\to \pi^0 K^{*\pm}$ & $ 6.9 \pm 2.3 $ & 6.5 &
4.0&6.8&6.2&7.1&4.5& $4.3^{+5.0\,(+1.7)}_{-2.2\,(-1.0)}$
\\
$B^0\to \pi^\mp K^{*\pm}$ & $ 9.8\pm 1.1 $ & 8.1 &
5.1&9.3&8.8&9.9&6.0& $6.0^{+6.8\,(+2.4)}_{-2.6\,(-1.3)}$
\\
$B^0\to \pi^0 K^{*0} $ & $ 1.7 \pm 0.8 $ & 2.5 &
1.5&3.5&3.3&3.8&2.0& $2.0^{+1.2\,(+0.9)}_{-0.6\,(-0.4)}$
\\
\hline $B^\pm\to \rho^\pm K^{0}$ & $ < 48 $ & 9.7 &
3.6&6.1&7.8&6.4&6.7& $8.7^{+6.8\,(+6.4)}_{-4.4\,(-4.3)}$
\\
$B^\pm\to \rho^0 K^{\pm}$ & $ 4.27^{+0.54}_{-0.56} $ & 4.3 &
2.5&4.0&5.0&4.3&3.8& $5.1^{+4.1\,(+3.6)}_{-2.8\,(-2.6)}$
\\
$B^0\to \rho^\mp K^{\pm}$ & $ 9.9^{+1.6}_{-1.5} $ & 10.1 &
4.7&7.0&7.8&7.2&7.9& $8.8^{+6.8\,(+6.2)}_{-4.5\,(-3.9)}$
\\
$B^0\to \rho^0 K^{0} $ & $ 5.6 \pm 1.1 $ &6.2 &
2.5&3.5&3.9&3.5&4.4& $4.8^{+4.3\,(+3.2)}_{-2.3\,(-2.0)}$
\\
\hline $B^\pm\to \omega K^{\pm} $ & $6.9\pm 0.5 $ & 5.9 &
2.1&5.9&9.3&6.4&4.8& $10.6^{+10.4\,(+7.2)}_{-5.8\,(-4.4)}$
\\
$B^0\to \omega K^{0} $ & $4.8\pm 0.6 $ & 4.9 &
1.9&6.0&8.6&6.6&4.8& $9.8^{+8.6\,(+6.7)}_{-4.9\,(-4.3)}$
\\
\hline\hline
\end{tabular}
\end{center}
\caption{Branching ratios in the NDR scheme in units of $10^{-6}$.
The label LO$_{\rm NLOWC}$ means the LO results with the NLO
Wilson coefficients, and +VC, +QL, +MP, and +NLO mean the
inclusions of the vertex corrections, of the quark loops, of the
magnetic penguin, and of all the above NLO corrections,
respectively. The errors in the parentheses arise only from the
variation of the hadronic parameters.} \label{pvbr}
\end{table}

The NLO PQCD results for the $B\to PV$ branching ratios are
displayed in Table~\ref{pvbr}, where the QCDF predictions from the
scenario S4 for the hadronic parameters \cite{BN} are shown for
comparison. It has been known that QCDF gives the branching ratios
similar to those in PQCD, if adopting S4. The QCDF predictions from
the default scenario \cite{BN} are usually smaller than the data by
a factor 2. It is found that the NLO Wilson evolution, labelled by
LO$_{\rm NLOWC}$, has a more significant effect on the $B\to K\phi$
decays than on the $B\to \pi K$ ones \cite{LMS05}. The reason is
that the former depends on the Wilson coefficient $a_5$ (see
Table~\ref{kphi}), which is enhanced more at NLO compared to $a_6$
in the latter. However, $a_5$ also receives a significant
destructive vertex correction as shown in Table~\ref{scale}, such
that the $B\to K\phi$ branching ratios drop rapidly from the
column LO$_{\rm NLOWC}$ to +VC in Table~\ref{pvbr}. The magnetic
penguin correction, labelled by +MP, further decreases the
branching ratios by about 30\%. Eventually, the $B\to K\phi$
branching ratios in NLO PQCD are still consistent with the
measured values.

For the $B\to\pi K^*$ branching ratios, our values in the column
LO$_{\rm NLOWC}$ are close to the LO PQCD predictions in
\cite{K02}. The effects from the vertex corrections, the quark
loops, and the magnetic penguin follow the pattern appearing in
the $B\to\pi K$ modes \cite{LMS05}. The main difference between
these two decays is that the former involve only $a_4$ (no $a_6$).
Therefore, the $B\to\pi K^*$ branching ratios are expected to be
much smaller than the $B\to\pi K$ ones. As indicated in
Table~\ref{pvbr}, our NLO PQCD predictions (central values) for
the former are about 1/3 of those for the latter. Confronting with
the data, the NLO predictions for the $B^\pm\to\pi^\pm K^{*0}$,
$B^\pm\to\pi^0 K^{*\pm}$, and $B^0\to\pi^\mp K^{*\pm}$ branching
ratios reach only 2/3 of the measured values. Nevertheless,
considering the uncertainties of both the theoretical predictions
and the experimental data, the discrepancy is not serious. On the
contrary, the PQCD results for the $B^0\to\pi^0 K^{*0}$ branching
ratios are in good agreement with the data.

Our LO PQCD results for the $B\to\rho K$ decays in
Table~\ref{pvbr} are close to those obtained in \cite{Chen01}:
$B(B^\pm\to \rho^\pm K^{0})=2.96\times 10^{-6}$, $B(B^\pm\to
\rho^0 K^{\pm})=2.18\times 10^{-6}$, $B(B^0\to \rho^\mp
K^{\pm})=5.42\times 10^{-6}$, and $B(B^0\to \rho^0
K^{0})=2.49\times 10^{-6}$. That is, the LO predictions fall short
by a factor 2 compared to the data \cite{HFAG}. The $B\to\rho K$
branching ratios, also smaller than the $B\to\pi K$ ones, are
attributed to the destructive combination of the Wilson
coefficients $a_4-2(m_{0K}/m_B)a_6$ in the former, and to the
constructive combination $a_4+2(m_{0K}/m_B)a_6$ in the latter [see
the replacement $m_{0K}\to -m_{0K}$ in Eq.~(\ref{re2})]. The NLO
Wilson evolution enhances the $B\to\rho K$ branching ratios, but
not sufficiently. Since the destructive combination flips sign of
the penguin contribution to the $B\to\rho K$ decays (comparing
Table~\ref{ta} here and Table~V in \cite{LMS05}), the NLO effects
change accordingly. For example, the magnetic-penguin correction
becomes constructive. The $B\to\rho K$ branching ratios then
increase all the way up to the measured values as shown in
Table~\ref{pvbr}. It turns out that the relation $B(B\to\rho
K)<B(B\to\pi K^*)$ in LO PQCD and in FA \cite{fac} is reversed
into $B(B\to\rho K)>B(B\to\pi K^*)$ at NLO.

Our LO PQCD results for the $B\to\omega K$ branching ratios are
also consistent with those in \cite{Chen01}: $B(B^\pm\to \omega
K^{\pm})=3.22\times 10^{-6}$ and $B(B^0\to \omega
K^{0})=2.07\times 10^{-6}$, which are smaller than the data
\cite{HFAG}. Similar to $B\to K\phi$, the $B\to\omega K$ decays
involve the Wilson coefficient $a_5$ through the color-suppressed
QCD penguin amplitude $F_{eK}^P$, in addition to $a_4$ and $a_6$
(see Table~\ref{rhok}). Hence, the enhancement of the $B\to\omega
K$ branching ratios from the NLO Wilson evolution is substantial.
Because of the sign flip of the QCD penguin contribution as in
$B\to\rho^0 K$, the vertex correction to $a_5$ becomes
constructive, opposite to that in the $B\to K\phi$ case (see the
opposite contributions of $P^{c\prime}$ in these two modes in
Table~\ref{ta}). This explains the jump of the $B\to\omega K$
branching ratios from the column LO$_{\rm NLOWC}$ to +VC in
Table~\ref{pvbr}. Though the magnetic penguin correction is
destructive, the net NLO predictions for the $B\to\omega K$
branching ratios remain large, with the central values being
higher than the observed ones.

As stated before, $a_5$ and its associated vertex correction
exhibit dramatic running effects at a low scale \cite{LMS05}.
Therefore, if lifting the hard characteristic scales in the
factorization formulas slightly (notice $\langle t\rangle\approx
1.0-1.5$ GeV for $P^{c\prime}$), the $a_5$ contribution will be
moderated. Then the predicted $B\to K\phi$ and $B\to\omega K$
branching ratios will increase and decrease, respectively,
approaching the central values of the data. However, we shall not
attempt such a fine tuning here, but point out that the comparison
of the two branching ratios provides an interesting test on the
significance of the $a_5$ contribution, i.e., on the sensitivity
of two-body nonleptonic $B$ meson decays to low-energy dynamics.
Note that the relevant form factors $A_0^{B\rho}$ and
$A_0^{B\omega}$ are roughly equal as indicated in
Eq.~(\ref{form}). If allowing the hard scale to be as low as 0.5
GeV, $B(B\to\omega K)$ would become three times of $B(B\to\rho^0
K)$. Because the measured branching ratios are close to each
other, $a_5$ plays a minor role, and a higher hard scale is
implied. It is then likely that two-body nonleptonic $B$ meson
decays are insensitive to low-energy dynamics, and a perturbation
theory is applicable to these processes.

\begin{table}[hbt]
\begin{center}
\begin{tabular}{cccccccccc}
\hline\hline Mode & Data~\cite{HFAG}& QCDF(S4)\,\cite{BN} & LO &
LO$_{\rm NLOWC}$& +VC & +QL &  +MP  & +NLO
\\
\hline $B^\pm\to K^\pm \phi $ & $3.7\pm 5.0$ &$0.7$&
$-2$&$-1$&$-1$&$0$&$-1$& $1^{+0\,(+0)}_{-1\,(-1)}$
\\
$B^0\to K^0 \phi$ & $9\pm 14$ &$0.8$& $0$&$0$&$0$&$1$&$0$&
$3^{+1\,(+0)}_{-2\,(-1)}$
\\
\hline $B^\pm\to \pi^\pm K^{*0}$ & $ -8.6 \pm  5.6$ &$0.8$&
$-3$&$-2$&$-2$&$0$&$-2$& $-1^{+1\,(+1)}_{-0\,(-0)}$
\\
$B^\pm\to \pi^0 K^{*\pm}$ & $ \phantom{-}4 \pm 29 $ &$-6.5$&
$-38$&$-31$&$-21$&$-29$&$-45$& $-32^{+21\,(+16)}_{-28\,(-19)}$
\\
$B^0\to \pi^\mp K^{*\pm}$ & $ -5 \pm 14 $ &$-12.1$&
$-56$&$-40$&$-42$&$-38$&$-60$& $-60^{+32\,(+20)}_{-19\,(-15)}$
\\
$B^0\to \pi^0 K^{*0} $ & $-1^{+27}_{-26} $ &$1.0$&
$-5$&$1$&$-13$&$2$&$3$& $-11^{+7\,(+5)}_{-5\,(-2)}$
\\
\hline $B^\pm\to \rho^\pm K^{0}$ & --- &$0.8$&
$2$&$1$&$1$&$1$&$1$& $1\pm 1\,(\pm 1)$
\\
$B^\pm\to \rho^0 K^{\pm}$ & $ 31^{+11}_{-10} $ &$31.7$&
$79$&$74$&$73$&$71$&$79$& $71^{+25\,(+17)}_{-35\,(-14)}$
\\
$B^0\to \rho^\mp K^{\pm}$ & $ 17^{+15}_{-16} $ &$20.0$&
$83$&$72$&$69$&$71$&$64$& $64^{+24\,(+7)}_{-30\,(-11)}$
\\
$B^0\to \rho^0 K^{0} $ & --- &$-2.8$& $7$&$-8$&$3$&$-8$&$-6$&
$7^{+8\,(+7)}_{-5\,(-4)}$
\\
\hline $B^\pm\to \omega K^{\pm} $ & $5\pm 6$ &$19.3$&
$82$&$45$&$37$&$43$&$57$& $32^{+15\,(+4)}_{-17\,(-5)}$
\\
$B^0\to \omega K^{0} $ & --- &$3.7$& $-4$&$7$&$-2$&$7$&$8$&
$-3^{+2\,(+2)}_{-4\,(-3)}$
\\
\hline\hline
\end{tabular}
\end{center}
\caption{Direct CP asymmetries in the NDR scheme in percentage.}
\label{pvcp}
\end{table}

The NLO PQCD predictions for the direct CP asymmetries of the
penguin-dominated $B\to PV$ decays are listed in Table~\ref{pvcp},
where the QCDF results from S4 \cite{BN} are also shown. The QCDF
results from the default scenario are always opposite in sign
compared with the data. Our predictions for the direct CP
asymmetries have larger magnitude here, since the updated CKM matrix
element $|V_{ub}|$ has increased substantially. If adopting $|V_{ub}|$
extracted from exclusive decays alone, and a lower $B$ meson decay
constant, the magnitude of the above direct CP asymmetries could
drop by 30\% easily. To be cautious, we emphasize only the various
patters of the direct CP asymmetries exhibited by the considered
$B\to PV$ decays, instead of their actual values. Because the NLO
Wilson evolution increases the penguin amplitudes, i.e., the $B\to
PV$ branching ratios, it dilutes the direct CP asymmetries. The
effects from the quark-loop and magnetic-penguin corrections can
be understood in the same way. The direct CP asymmetries of the
$B\to K\phi$ modes, which are almost pure-penguin processes,
remain vanishing at NLO as shown in Table~\ref{pvcp}.

We explain the different effects from the vertex corrections on
the direct CP asymmetries $A_{CP}(B^\pm \to \pi^0 K^{*\pm})$,
$A_{CP}(B^\pm \to \rho^0 K^\pm)$, and $A_{CP}(B^\pm \to \pi^0
K^\pm)$. It has been known \cite{LMS05} that the penguin amplitude
$P'$ in the $B\to\pi K$ decays is in the second quadrant, with a
strong phase about $15^o$ relative to the negative real axis, and
that the source of this phase arises mainly from the almost
imaginary scalar penguin annihilation. The color-allowed tree
amplitude $T'$ is roughly aligned with the positive real axis.
Consequently, the $B^0\to\pi^\mp K^\pm$ decays, involving $P'$ and
$T'$, show a sizable direct CP asymmetry \cite{KLS}. On the other
hand, the color-suppressed tree amplitude $C'$, enhanced by the
vertex corrections, becomes almost imaginary. It then orients the
sum $T'+C'$ into the fourth quadrant, such that $T'+C'$ and
$P'+P'_{ew}$ more or less line up (point to the opposite
directions) in the $B^\pm \to \pi^0 K^\pm$ decays. This is the
reason the magnitude of $A_{CP}(B^\pm \to \pi^0 K^\pm)$ tends to
vanish in NLO PQCD, leading to the pattern \cite{LMS05},
\begin{eqnarray}
|A_{CP}(B^\pm \to \pi^\mp K^\pm)|\gg |A_{CP}(B^\pm \to \pi^0
K^\pm)|\;,\label{dpik}
\end{eqnarray}
which is in agreement with the data.

The pattern of the direct CP asymmetries in the $B\to\pi K^*$
decays can be described similarly, but with the distinction that
their penguin emission amplitudes contain only $a_4$ (no $a_6$).
The almost imaginary scalar penguin annihilation then renders $P'$
more inclined to the positive imaginary axis compared to that in
the $B\to\pi K$ case. Hence, $T'$ and $P'$ are more orthogonal to
each other in the $B^0 \to \pi^\mp K^{*\pm}$ decays. For this
configuration, the modification from $C'$ will be weaker. That is,
the total tree amplitude $T'+C'$ and the total penguin amplitude
$P'+P'_{ew}$ will not line up completely in the $B^\pm \to \pi^0
K^{*\pm}$ modes. We then predict the larger $B\to\pi K^*$ direct
CP asymmetries, and the pattern shown in Table~\ref{pvcp},
\begin{eqnarray}
|A_{CP}(B^\pm \to \pi^\mp K^{*\pm})|>|A_{CP}(B^\pm \to \pi^0
K^{*\pm})|\;.\label{dpiks}
\end{eqnarray}

The situation in the $B\to\rho K$ decays is even more extreme,
where the real part of $P'$ almost diminishes due to the
destructive combination of the Wilson coefficients
$a_4-2(m_{0K}/m_B)a_6$. Because of the sign flip of the penguin
annihilation amplitude $F_{a}^P$ in Table~\ref{rhok} under the
replacements $m_{0\pi}\to -m_\rho$ and $m_{0K}\to -m_{0K}$ in
Eq.~(\ref{re2}), $P'$ is roughly aligned with the negative
imaginary axis as indicated in Table~\ref{ta}. This explains the
predicted sign of $A_{CP}(B^0 \to \rho^\mp K^\pm)$ opposite to
that of $A_{CP}(B^0 \to \pi^\mp K^\pm)$. In this case $T'$ and
$P'$ have the maximal relative strong phase, such that the
modification from $C'$ is not obvious. We then predict the larger
$B\to\rho K$ direct CP asymmetries, and the pattern,
\begin{eqnarray}
A_{CP}(B^\pm \to \rho^\mp K^\pm)\approx A_{CP}(B^\pm \to \rho^0
K^\pm)\;.\label{drhok}
\end{eqnarray}
That is, the NLO corrections to the $B\to\rho K$ direct CP
asymmetries are minor, and the LO predictions derived in
\cite{Chen01} are reliable.

The direct CP asymmetries in the $B\to\omega K$ decays can be
understood by means of those in $B\to\rho K$ plus the effect from
the color-suppressed QCD penguin amplitude $P^{c\prime}$. Since
$P^{c\prime}$ is roughly aligned with the positive real axis as
shown in Table~\ref{ta}, the relative strong phase between the
tree and penguin amplitudes decreases from the maximal
configuration in the $B\to\rho K$ decays. Therefore, the vertex
corrections have some effect on $A_{CP}(B^\pm \to \omega K^\pm)$,
which is then expected to be smaller than $A_{CP}(B^\pm\to\rho^0
K^\pm)$, but still sizable as indicated in Table~\ref{pvcp}. We
stress that the above predictions for the direct CP asymmetries,
if confirmed by the future data, will support the source of strong
phases from the scalar penguin annihilation in PQCD.

\begin{table}[hbt]
\begin{center}
\begin{tabular}{cccccccc}
\hline\hline  & Data \cite{HFAG}&
 LO & LO$_{\rm NLOWC}$ & +VC
& +QL &  +MP  & +NLO
\\
\hline
$S_{\phi K_S}$ &$ 0.47\pm 0.19$ &
$0.72$&$0.72$&$0.72$&$0.72$&$0.72$&
$0.71^{+0.01\,(+0.01)}_{-0.01\,(-0.00)}$\\
$S_{\rho^0 K_S}$ & $0.17 \pm 0.58$ &
$0.71$&$0.66$&$0.47$&$0.67$&$0.66$&
$0.50^{+0.10\,(+0.04)}_{-0.06\,(-0.04)}$\\
$S_{\omega K_S}$ & $0.64\pm 0.30$ &
$0.76$&$0.69$&$0.84$&$0.69$&$0.70$&
$0.84^{+0.03\,(+0.00)}_{-0.07\,(-0.02)}$\\
\hline\hline
\end{tabular}
\end{center}
\caption{Mixing-induced CP asymmetries.
}\label{mixcp}
\end{table}

The NLO PQCD predictions for the mixing-induced CP asymmetries
$S_{MK_S}$, $M=\phi$, $\rho^0$, and $\omega$, are collected in
Table~\ref{mixcp}. It is found that all $S_{MK_S}$ exhibit
positive deviations from those of the tree-dominated $b\to c\bar
cs$ transitions at LO. The deviation in the $B^0\to\phi K_S$ decay
remains negligible at NLO due to the absence of the tree pollution
from the color-suppressed tree amplitude. It is interesting to see
that the NLO corrections turn $S_{\rho^0K_S}$ into a much lower
value about 0.5. The reason is, again, the destructive combination
of the Wilson coefficients $a_4-2(m_{0K}/m_B)a_6$, which flips the
sign of the penguin amplitude $P'$ as shown in Table~\ref{ta}. The
color-suppressed tree amplitude $C'$, modified by the vertex
corrections, then becomes parallel to $P'$, leading to the very
negative deviation. This observation is consistent with the
tendency indicated by the data. The deviation of $S_{\omega K_S}$
is positive, since the sign of $P'$ is opposite to that in the
$B^0\to\rho^0 K_S$ decay as shown in Tables~\ref{rhok} and
\ref{ta}. The large deviations exhibited by the $B^0\to\rho^0
K_S$, $\omega K_S$ decays at NLO are attributed to the involved
large $C'$ (see Table~\ref{ta}). Our predictions are basically in
agreement with those in QCDF \cite{B05,BHNR}, but differ from
those in QCDF plus FSI \cite{CCS2}, which predicted
$S_{\rho^0K_S}\approx 0.76$ with a positive deviation.

\section{CONCLUSION}

In this paper we have surveyed the penguin-dominated $B\to PV$
decays in the NLO PQCD formalism, concentrating on the $B\to
K\phi$, $\pi K^*$, $\rho K$, and $\omega K$ modes. The NLO Wilson
evolution enhances all the branching ratios, while the other NLO
corrections modulate the branching ratio of each mode in a
different way. For example, the vertex corrections to the Wilson
coefficient $a_5$ decrease (increase) the $B\to K\phi$
($B\to\omega K$) branching ratios significantly. The net NLO
effects modify the LO predictions for the $B\to K\phi$, $\pi K^*$
branching ratios a bit, but increase those for the $B\to\rho K$,
$\omega K$ ones by a factor more than 2. Owing to this dramatic
enhancement, the NLO PQCD results for these $B\to PV$ branching
ratios are in better agreement with the data. Since the penguin
emission amplitudes in the $B\to\pi K$, $\pi K^*$, and $\rho K$
decays are proportional to $a_4+2(m_{0K}/m_B)a_6$, $a_4$, and
$a_4-2(m_{0K}/m_B)a_6$, respectively, their real parts decrease in
the above order. The almost imaginary scalar penguin annihilation
then makes the total penguin amplitudes more orthogonal to the
color-allowed tree amplitudes. As a consequence, the modifications
from the NLO color-suppressed tree amplitudes decrease in the
above order. We then predict the patterns of the direct CP
asymmetries in Eqs.~(\ref{dpik})--(\ref{drhok}). These different
patters, if confirmed by the data, will support the source of
strong phases from the scalar penguin annihilation in PQCD. We
emphasize that PQCD is the only theoretical approach available in
the literature, which can produce the above patterns of the direct
CP asymmetries.

A remark is in order. As mentioned in the Introduction, $B$ meson
transition form factors are not factorizable in collinear
factorization theorem due to the existence of the end-point
singularities. Recently, a zero-bin regularization for these
singularities have been proposed, such that $B$ meson transition
form factors become factorizable \cite{MS0605}. The idea is to
subtract soft modes from collinear effective fields to avoid
double counting of soft dynamics. When implemented in a
factorization formula, this regularization scheme is equivalent to
cutoffs at small momentum fractions $x$. The zero-bin
regularization has been extended to annihilation amplitudes in
two-body nonleptonic $B$ meson decays, which then also become
factorizable \cite{ALRS}. As expected, the annihilation amplitudes
are real, because internal particles, without carrying parton
transverse momenta, do not go on mass shell, when one introduces
cutoffs at small $x$. Therefore, whether a physical strong phase
is generated by the LO $B$ meson annihilation amplitude is a
distinction between this modified collinear factorization theorem
and the $k_T$ factorization theorem.

We have also predicted the mixing-induced CP asymmetries of the
penguin-dominated $B\to PV$ decays. Among the decays we have
investigated ($B^0\to\pi^0 K_S$, $\phi K_S$, $\rho^0 K_S$, $\omega
K_S$), $B^0\to\phi K_S$ is the cleanest mode for extracting the
standard-model parameter $\sin(2\phi_1)$, since it does not
involve the color-suppressed tree amplitude, which can be greatly
enhanced by the NLO effects. It has been found that $S_{\rho^0
K_S}$ exhibits the maximal negative deviation from those of the
tree-dominated $b\to c\bar cs$ transitions, and all other $S_f$
show positive deviations in NLO PQCD. Hence, it is still puzzling,
when confronting the theoretical predictions, especially those for
$S_{\phi K_S}$ and $S_{\pi^0 K_S}$, with the experimental
observations.

\vskip 1.0cm

We thank C.H. Chen, H.Y. Cheng, C.K. Chua, W.S. Hou, C.M. Jen, M.
Nagashima, and A. Soni for useful discussions. This work was
supported by the National Science Council of R.O.C. under Grant
No. NSC-94-2112-M-001-001, by the Taipei branch of the National
Center for Theoretical Sciences, and by the U.S. Department of
Energy under Grant No. DE-FG02-90ER40542. HNL thanks the Yukawa
Institute for Theoretical Physics, Kyoto University, for her
hospitality during his visit.

\appendix

\section{DISTRIBUTION AMPLITUDES}

The pion and kaon distribution amplitudes \cite{BBL0603}, and the
$\rho$ and $K^*$ meson distribution amplitudes \cite{BZ0603} have
been updated up to two-parton twist-3 level in QCD sum rules. The
twist-2 pion (kaon) distribution amplitude $\phi^A_{\pi(K)}$ and
the twist-3 ones $\phi_{\pi(K)}^P$ and $\phi_{\pi(K)}^T$ have been
parameterized as \cite{BBL0603}
\begin{eqnarray}
\phi_{\pi(K)}^A(x) &=& \frac{f_{\pi(K)}}{2\sqrt{2N_c}}\, 6x(1-x)
\left[1 + a_1^{\pi(K)} C_1^{3/2}(2x-1) +
a_2^{\pi(K)}C_2^{3/2}(2x-1)
\right] \;,
\end{eqnarray}
and
\begin{eqnarray}
\phi^P_{\pi(K)}(x) &=& \frac{f_{\pi(K)}}{2\sqrt{2N_c}}
\bigg[
1 + 3\rho^{\pi(K)}_+ \left(1+6 a^{\pi(K)}_2\right)
- 9 \rho^{\pi(K)}_- a^{\pi(K)}_1
\nonumber\\
&&
\hspace{10mm}
+ C_1^{1/2}(2x-1) \left[\frac{27}{2}\,
\rho^{\pi(K)}_+ a^{\pi(K)}_1 - \rho^{\pi(K)}_-
\left( \frac{3}{2} + 27 a^{\pi(K)}_2\right)\right]
\nonumber\\
&&
\hspace{10mm}
 + C_2^{1/2}(2x-1)
\left( 30 \eta_{3\pi(K)} + 15 \rho^{\pi(K)}_+ a^{\pi(K)}_2
- 3 \rho^{\pi(K)}_- a^{\pi(K)}_1\right)
\nonumber\\
&&
\hspace{10mm}
+ C_3^{1/2}(2x-1)
\left( 10 \eta_{3\pi(K)} \lambda_{3\pi(K)}
- \frac{9}{2}\,\rho^{\pi(K)}_- a^{\pi(K)}_2\right)
 - 3\eta_{3\pi(K)} \omega_{3\pi(K)} C_4^{1/2}(2x-1)
\nonumber\\
&&\hspace{10mm}
 + \frac{3}{2}\,\left(\rho^{\pi(K)}_+ + \rho^{\pi(K)}_-\right)
\left(1-3a^{\pi(K)}_1+6a^{\pi(K)}_2\right)\ln x
\nonumber\\
&&\hspace{10mm}
 + \frac{3}{2}\,\left(\rho^{\pi(K)}_+-\rho^{\pi(K)}_-\right)
\left(1+3 a^{\pi(K)}_1 + 6  a^{\pi(K)}_2\right) \ln \bar x
\bigg]\;,
\\
\phi^T_{\pi(K)}(x) &=& \frac{f_{\pi(K)}}{2\sqrt{2N_c}}
\bigg\{
(1-2x)
\bigg[
1 + \frac{3}{2}\,\rho^{\pi(K)}_+ + 15
\rho^{\pi(K)}_+ a^{\pi(K)}_2
- \frac{15}{2}\,\rho^{\pi(K)}_- a^{\pi(K)}_1
\nonumber\\
&&
\hspace{30mm}
+ \left( 3 \rho^{\pi(K)}_+ a^{\pi(K)}_1
- \frac{15}{2}\,\rho^{\pi(K)}_- a^{\pi(K)}_2\right) 3(2x-1)
\nonumber\\
&&
\hspace{30mm}
+
\left( 5\eta_{3\pi(K)} -\frac{1}{2}\,\eta_{3\pi(K)}\omega_{3\pi(K)}
+ \frac{3}{2}\,\rho^{\pi(K)}_+ a^{\pi(K)}_2 \right)
6(1-10x+10x^2)
\nonumber\\
&&
\hspace{30mm}
 +\,
5\eta_{3\pi(K)} \lambda_{3\pi(K)}
(1-2x)
\left[ 35x(1-x)-2 \right]
\nonumber\\
&&
\hspace{30mm}
+
\frac{3}{2} \left(\rho^{\pi(K)}_++\rho^{\pi(K)}_-\right)
\left(1-3 a^{\pi(K)}_1+6 a^{\pi(K)}_2\right)
\ln x
\nonumber\\
&&
\hspace{30mm}
+
\frac{3}{2} \left(\rho^{\pi(K)}_+-\rho^{\pi(K)}_-\right)
\left(1+3 a^{\pi(K)}_1+6 a^{\pi(K)}_2\right)
\ln (1-x)
\bigg]
\nonumber\\
&&
\hspace{10mm}
+ \left( 3 \rho^{\pi(K)}_+ a^{\pi(K)}_1 -
\frac{15}{2}\,\rho^{\pi(K)}_- a^{\pi(K)}_2\right) 6x(1-x)
- 15 \eta_{3\pi(K)} \lambda_{3\pi(K)} x(1-x)
\nonumber\\
&&
\hspace{10mm}
+
\frac{3}{2} \left(\rho^{\pi(K)}_++\rho^{\pi(K)}_-\right)
\left(1-3 a^{\pi(K)}_1+6 a^{\pi(K)}_2\right)
(1-x)
\nonumber\\
&&
\hspace{10mm}
-
\frac{3}{2} \left(\rho^{\pi(K)}_+-\rho^{\pi(K)}_-\right)
\left(1+3 a^{\pi(K)}_1+6 a^{\pi(K)}_2\right)
x
\bigg\}\;,
\end{eqnarray}
respectively, with $a_1^\pi=0$, $a_1^K=0.06$, $a_2^\pi=a_2^K=0.25$ and
\begin{eqnarray}
\eta_{3\pi(K)} &=& \frac{f_{3\pi(K)}}{f_{\pi(K)}}\,
\frac{m_{q(s)}+m_q}{m_{\pi(K)}^2} \, =\,
\frac{f_{3\pi(K)}}{f_{\pi(K)}\, m_{0\pi(K)}}
\;,\\
\rho_{+}^{\pi(K)} &=& \frac{(m_{q(s)}+m_q)^2}{m^2_{\pi(K)}} \, =\,
\frac{m_{\pi(K)}^2}{m_{0\pi(K)}^2}
\;,\\
\rho_{-}^{\pi(K)} &=& \frac{m_{q(s)}^2-m_q^2}{m_{\pi(K)}^2} \;,
\end{eqnarray}
where $m_q$ is the mass of the light quarks $u$ and $d$. One has
$\rho_{+}^{\bar K} =  \rho_{+}^K$ and $\rho_{-}^{\bar K} =
-\rho_{-}^K$ for $\bar K$, and numerically $\rho_{+}^K \simeq
\rho_{-}^K$ and $\rho_{-}^\pi\simeq 0$. The explicit values of the
other relevant parameters are $f_{3\pi}=f_{3K}=0.0045$ GeV$^2$,
$\lambda_{3\pi}=0.0$, $\lambda_{3K}=1.6$, $\omega_{3\pi}=-1.5$,
and $\omega_{3K}=-1.2$
 \cite{BBL0603}. In the above kaon distribution amplitudes the
momentum fraction $x$ is carried by the $s$ quark \footnote{The
$K$ ($\bar K$) meson corresponds to the $K^-$ and $\bar K^0$
($K^+$ and $K^0$) mesons in \cite{BBL0603}. The $\bar K$ meson
distribution amplitudes are given by $\phi_{\bar K}(x) =
\phi_K(1-x)$ or $\phi_{\bar K}(x) = \phi_K(x)$ but with a sign
flip of $a_1^{\bar K}$, $\rho_{-}^{\bar K}$, and $\lambda_{3\bar
K}$.}. The Gegenbauer polynomials are defined by
\begin{eqnarray}
C_0^\lambda(t) &=& 1
\;,\\
C_1^\lambda(t) &=& 2\lambda t
\;,\\
C_2^\lambda(t) &=& 2\lambda (\lambda+1)t^2 -\lambda
\;,\\
C_3^\lambda(t) &=&
\frac{4\lambda}{3}(\lambda^2+3\lambda+2)t^3-2\lambda(\lambda+1)t
\;,\\
C_4^\lambda(t) &=&
\frac{2\lambda}{3}(\lambda^3+6\lambda^2+11\lambda+6)t^4
-2\lambda(\lambda^2+3\lambda+2)t^2+\frac{\lambda}{2}(\lambda+1)
\;.
\end{eqnarray}

The twist-2 distribution amplitudes for longitudinally polarized
vector mesons are parameterized as \cite{BZ0603}
\begin{eqnarray}
\phi_\rho(x)&=&\frac{3f_\rho}{\sqrt{2N_c}} x(1-x)\left[1+
a_{2\rho}^\parallel\, C_2^{3/2}(2x-1)\right]\;,
\label{pwr}\\
\phi_\omega(x)&=&\frac{3f_\omega}{\sqrt{2N_c}} x(1-x)\left[1+
a_{2\omega}^\parallel\, C_2^{3/2}(2x-1)\right]\;,
\label{pwomega}\\
\phi_{K^*}(x)&=&\frac{3f_{K^*}}{\sqrt{2N_c}}
x(1-x)\left[1+a_{1K^*}^\parallel\, C_1^{3/2}(2x-1)+
a_{2K^*}^\parallel\, C_2^{3/2}(2x-1)\right]\;,
\label{pwrk}\\
\phi _{\phi }\left( x\right)
&=&\frac{3f_\phi}{\sqrt{2N_{c}}}x(1-x)\left[1+
a_{2\phi}^\parallel\, C_2^{3/2}(2x-1)\right]\;. \label{phi2}
\end{eqnarray}
For the twist-3 vector meson distribution amplitudes $\phi^s$ and
$\phi^t$, we adopt their asymptotic models \cite{Li04}. Since the
Gegenbauer coefficients of these twist-3 distribution amplitudes
have not yet been well constrained, the asymptotic models are
allowed. Moreover, we shall vary the Gegenbauer coefficients of
the twist-2 distribution amplitudes by 100\%, which is larger than
the error specified in \cite{BZ0603}. Therefore, the theoretical
uncertainty of our predictions from this source is representative
enough.

\end{document}